\documentclass{conm-p-l}

\copyrightinfo{2010}{}

\usepackage{graphicx}
\usepackage{subfigure}

\usepackage{amssymb,amsmath,amsthm,amscd}

\theoremstyle{definition}

\theoremstyle{remark}

\numberwithin{equation}{section}

\newcommand{\Dl}{\Delta}

\newcommand{\hv}{\hat{v}}

\newcommand{\lag}{\langle}
\newcommand{\rag}{\rangle}

\begin{document}

\title[On the Arrow of Time]{On the Arrow of Time}

\author{Y. Charles Li}
\address{Department of Mathematics, University of Missouri, 
Columbia, MO 65211, USA}
\email{liyan@missouri.edu}

\author{Hong Yang}
\address{Mathematics of Networks and Communications Research Department,
Bell Laboratories, 600 Mountain Avenue, Murray Hill, NJ 07974, USA}
\email{h.yang@research.bell-labs.com}

\curraddr{}
\thanks{}

\subjclass{Primary 82; Secondary 37}
\date{}

\dedicatory{}


\begin{abstract}
We believe the following three ingredients are enough to explain the mystery of the arrow of time: (1). equations 
of dynamics of gas molecules, (2). chaotic instabilities of the equations of dynamics, (3). unavoidable 
perturbations to the gas molecules. The level of physical rigor or mathematical rigor that can be reached 
for such a theory is unclear.
\end{abstract}

\maketitle

\section{The Theory}

There have been a lot of writings on the arrow of time; nevertheless, the problem still remains a mystery. In the 
equilibrium thermodynamics context, the arrow of time refers to the second law of thermodynamics. Consider the 
example of gas in a box, each gas molecule moves according to the equation of dynamics (free flying and collision) 
which is reversible in the coordinate time; on the other hand, the whole body of gas can only move in an entropy 
increasing direction among thermodynamic equilibria, e.g. moving from half box to whole box. The entropy 
increasing direction can be viewed as an intrinsic time direction --- the arrow of time. In fact, there are many 
similar irreversible features in nature; for instances, living things can only grow older, sugar and water do not 
spontaneously un-mix, broken glass does not spontaneously fix itself. Explaining all these arrows of time is a 
challenging task. According to L. Boltzmann \cite{Bol74}:``The second law (of thermodynamics) can never be 
proved mathematically by means of the equations of dynamics alone." Here we shall approach the problem with 
the following three ingredients:
\begin{enumerate}
\item equations of dynamics of gas molecules, 
\item chaotic instabilities of the equations of dynamics, 
\item unavoidable perturbations to the gas molecules. 
\end{enumerate}

Trying to use chaos to explain irreversibility is not a new attempt, see for example I. Prigogine's writings \cite{Pri80}
\cite{PS84} \cite{PS88} \cite{Pri95}. Of course, there are many other attempts from different perspectives. The most 
recent one is that of Maccone \cite{Mac09} from quantum perspective, see also its critiques \cite{JR10} \cite{KL09} 
\cite{Nik09} \cite{Mac09b}.

\begin{figure}[ht] 
\centering
\includegraphics[width=4.5in,height=1.5in]{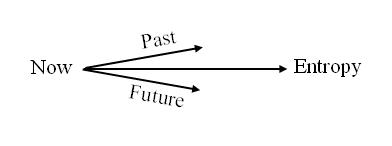}
\caption{The diagram of the arrow of time. The `Past', `Now' and `Future' are coordinate time, and 
the `Entropy' is the thermodynamic equilibrium entropy.}
\label{DA}
\end{figure}

As Boltzmann said, the equations of dynamics alone will not be able to explain irreversibility. Unavoidable 
perturbations to the gas molecules have to be taken into account. These perturbations are inside the isolated system, 
and they can be for examples physical noises and chemical fluctuations. If the dynamics of the gas 
molecules is non-chaotic, then these small perturbations will 
not generate any substantial effect; and the dynamics of the whole gas body will be reversible at least partially. 
The reality is 
that the system of equations of gas molecules is a non-integrable system and its dynamics is chaotic. Under 
such chaotic instabilities, the small unavoidable perturbations totally alter and liberate the dynamics of the 
whole gas body during the relaxation time from one thermodynamic equilibrium to another. In fact, the final 
thermodynamic equilibrium is independent of the initial conditions of the gas molecules, under such 
chaotic instabilities and unavoidable perturbations. For the example of gas in half box moving to full box, 
even the coordinate time and velocity of every gas molecule are reversed at any time during the relaxation 
time from half box to full box, the final thermodynamic equilibrium will still be the full box. Thus we arrive at 
the diagram for the arrow of time, Figure \ref{DA}. 

Starting from the coordinate time `Now', in both forward and backward coordinate time, the thermodynamic equilibrium
entropy is increasing. The entropy time is folded into only positive direction --- the arrow of time. In a more generalized 
sense, for a living thing at `Now', even its every molecule's coordinate time is reversed, the living thing will not get 
younger, rather continue to grow older!

Next we will explain a little more on chaotic instabilities and the effect of unavoidable perturbations. The main 
character of chaotic instabilities is `sensitive dependence on initial data'. This is commonly characterized by 
amplifications of initial perturbations. For the traditional lower dimensional chaos, such amplifications are 
exponential in time. For the dynamics of gas moleculesm such amplifications are actually linear in time as 
shown later in the numerical simulation of one dimensional gas dynamics. The dynamics of the gas 
molecules is chaotic in both forward and backward coordinate time. Due to the unavoidable perturbations, the real
physical orbits of the gas molecules are not the mathematical orbits prescribed by the equations of dynamics and the 
initial conditions. Only between two consecutive perturbations among all gas molecules, the gas molecules follow 
a segment of the mathematical orbits. The duration between two consecutive perturbations is extremely small. So 
the real physical orbits quickly diverge from the mathematical ones. For an $N$ particles system on the interval 
[$0,1$] for example, the phase space 
has $2N$ dimensions, coordinated by the particles' locations and velocities. The region where all particles are in 
the half interval [$0,1/2$], has measure $2^{-N}$. The main theme of our study is to investigate the possibility of return
to this small region when the time and velocities are reversed at some moment on the orbit originating from a point 
in this small region. 

\section{A Popular Challenger to Irreversibility --- The Poincar\'e Recurrence Theorem}

The unwarranted argument here is that due to the Poincar\'e recurrence theorem, full box gas will return near 
half box gas after enough time. Physicists making such an argument apply the Poincar\'e recurrence theorem 
without its condition. The crucial condition is that one can find a bounded invariant region in the phase space of 
all the gas molecules coordinated by their locations and momenta. Then inside the bounded invariant region, 
Poincar\'e recurrence theorem holds. The system of equations of gas molecules is a non-integrable system 
for which such a bounded invariant region does not exist. Even for near-integrable Hamiltonian systems, 
the well-known mechanism of Arnold diffusion can lead to unbounded drifting of the momenta. 

\section{Ergodicity}

There are two types of chaotic dynamics of gas molecules when the whole system moves from one 
thermodynamic equilibrium to another. Type I chaotic dynamics is the `transitional chaotic dynamics' during 
the relaxation time, while type II chaotic dynamics is the `saturated chaotic dynamics' when the system is 
in a thermodynamic equilibrium. The transitional chaotic dynamics has a non-zero mean motion, and it is the 
non-zero mean motion that eliminates the possibility of ergodicity. The saturated chaotic dynamics has no 
mean motion and may have ergodicity. Since the system is in a thermodynamic equilibrium, its entropy 
maintains constant and is compatible with ergodicity of the gas molecules' dynamics. 

\section{Numerical Simulations}

\subsection{1-Dimensional Dynamics}

\begin{figure}[ht] 
\centering
\includegraphics[width=4.5in,height=4.5in]{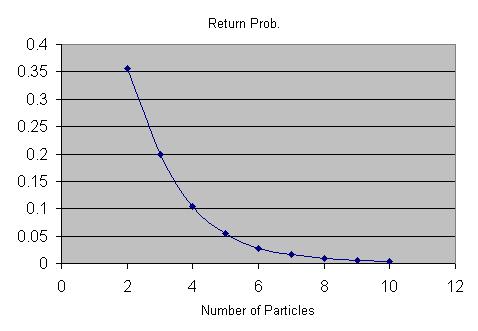}
\caption{Return probability for particles initially randomly positioned in the half interval [$0,0.5$].}
\label{1DF1}
\end{figure}

\begin{figure}[ht] 
\centering
\includegraphics[width=4.5in,height=4.5in]{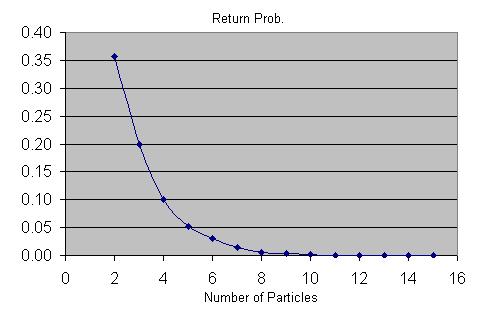}
\caption{Return probability for particles initially randomly positioned in the full interval [$0,1$].}
\label{1DF2}
\end{figure}

\begin{figure}[ht] 
\centering
\includegraphics[width=4.5in,height=4.5in]{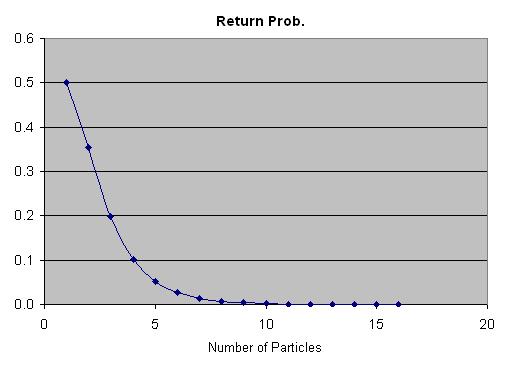}
\caption{Return probability for particles initially evenly positioned in the full interval [$0,1$].}
\label{1DF3}
\end{figure}

\begin{figure}[ht] 
\centering
\includegraphics[width=4.5in,height=4.5in]{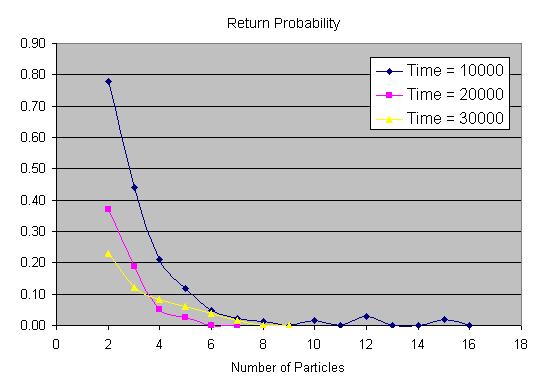}
\caption{Return (reverse) probability for different relaxation time.}
\label{1DF4}
\end{figure}

\begin{figure}[ht] 
\centering
\subfigure{\includegraphics[width=2.3in,height=2.3in]{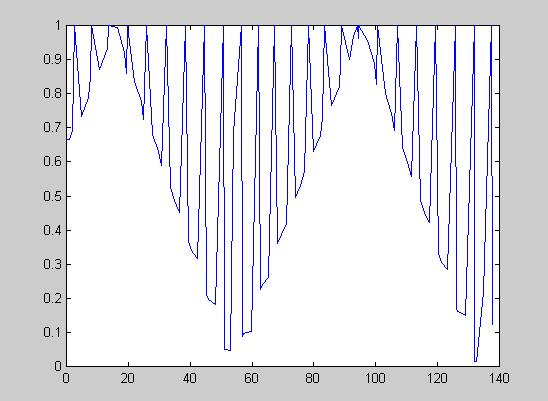}}
\subfigure{\includegraphics[width=2.3in,height=2.3in]{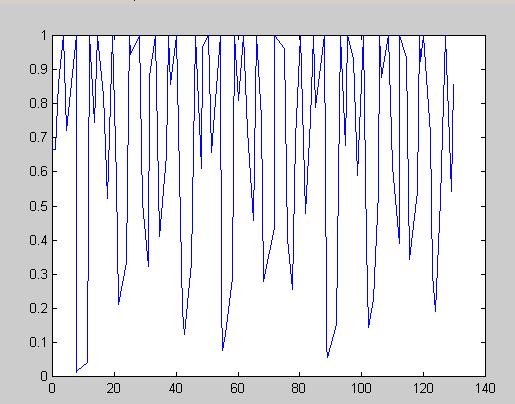}}
\subfigure{\includegraphics[width=2.3in,height=2.3in]{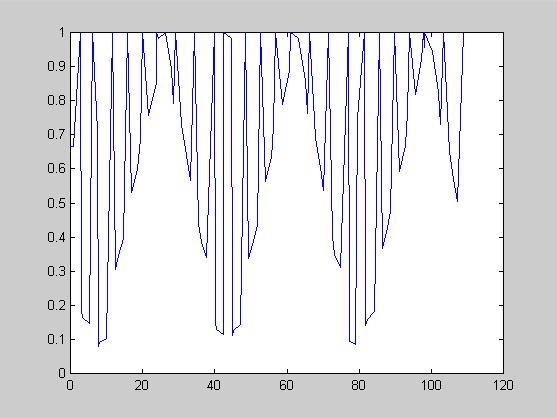}}
\subfigure{\includegraphics[width=2.3in,height=2.3in]{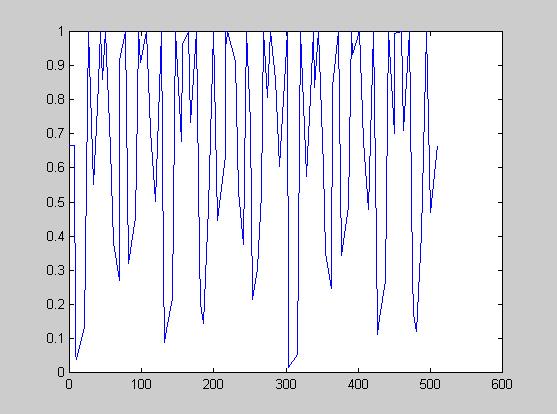}}
\caption{The location time series of the right particle in the two particles system.}
\label{fts}
\end{figure}

\begin{figure}[ht] 
\centering
\includegraphics[width=5.0in,height=4.0in]{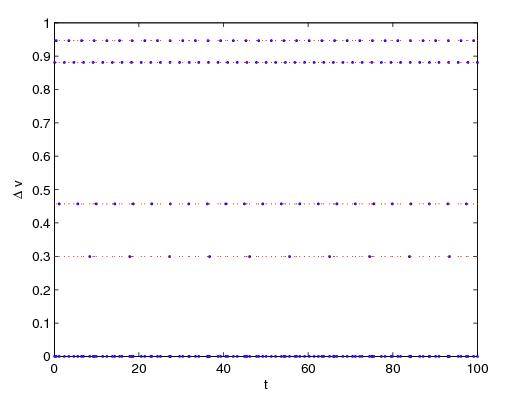}
\caption{Velocity deviation in Euclidean norm. $N_p = 2$. The larger dots mark the velocity deviation, and the 
smaller dots mark the exact horizontal lines. Initial perturbations are on both velocity and position. 
Each perturbation is bounded by $10^{-6}$.}
\label{fvd1}
\end{figure}

\begin{figure}[ht] 
\centering
\includegraphics[width=4.5in,height=4.5in]{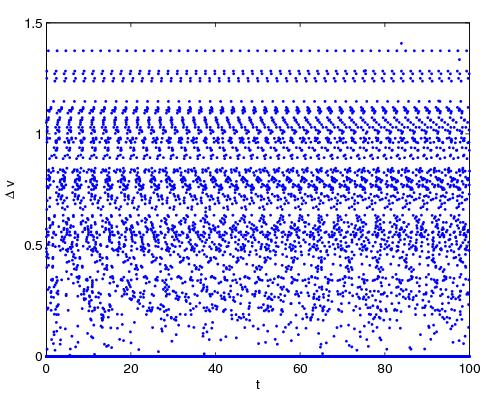}
\caption{Velocity deviation in Euclidean norm. $N_p = 16$. Initial perturbations are on both velocity and position. 
Each perturbation is bounded by $10^{-6}$.}
\label{fvd2}
\end{figure}

\begin{figure}[ht] 
\centering
\includegraphics[width=4.5in,height=4.0in]{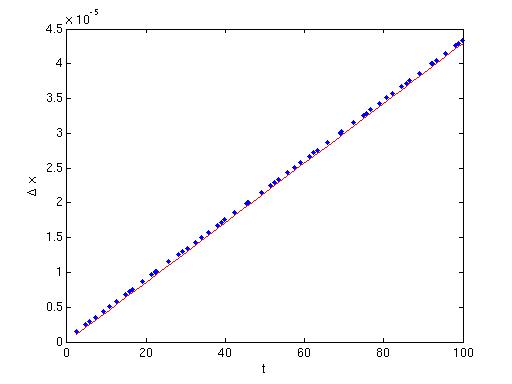}
\caption{Location deviation in Euclidean norm. $N_p = 2$. Initial perturbations are on both velocity and position. 
Each perturbation is bounded by $10^{-6}$.}
\label{fld1}
\end{figure}

\begin{figure}[ht] 
\centering
\includegraphics[width=5.0in,height=4.5in]{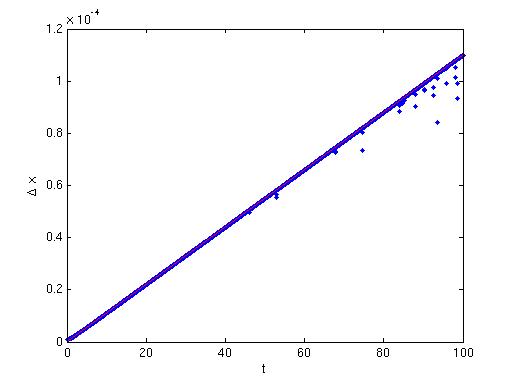}
\caption{Location deviation in Euclidean norm. $N_p = 16$. Initial perturbations are on both velocity and position. 
Each perturbation is bounded by $10^{-6}$.}
\label{fld2}
\end{figure}

\begin{figure}[ht] 
\centering
\includegraphics[width=3.0in,height=1.5in]{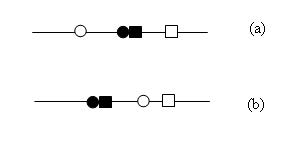}
\caption{Essential types of collisions.}
\label{2CD}
\end{figure}

\begin{figure}[ht] 
\centering
\includegraphics[width=4.5in,height=4.0in]{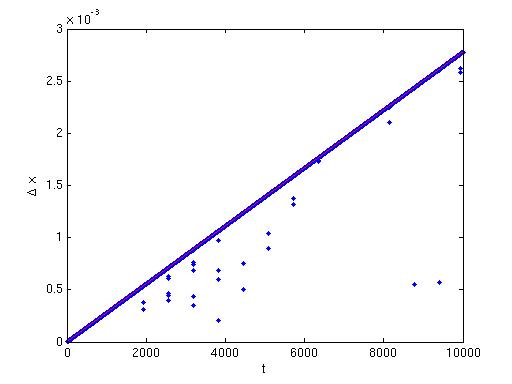}
\caption{Location deviation in Euclidean norm. $N_p = 2$. Initial perturbations are on both velocity and position. 
Each perturbation is bounded by $10^{-6}$.}
\label{fld3}
\end{figure}

\begin{figure}[ht] 
\centering
\includegraphics[width=4.5in,height=4.0in]{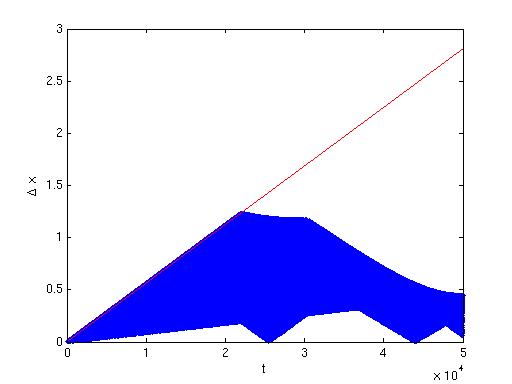}
\caption{Location deviation in Euclidean norm. $N_p = 2$. Initial perturbations are on both velocity and position. 
Each perturbation is bounded by $10^{-4}$.}
\label{fld4}
\end{figure}

\begin{figure}[ht] 
\centering
\includegraphics[width=4.5in,height=4.0in]{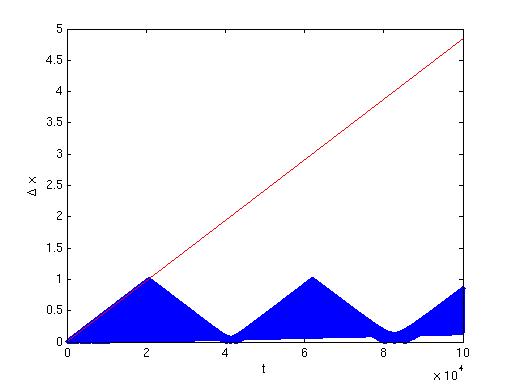}
\caption{Location deviation in Euclidean norm. $N_p = 2$. Initial perturbations are on both velocity and position. 
Each perturbation is bounded by $10^{-4}$.}
\label{fld5}
\end{figure}

\begin{figure}[ht] 
\centering
\includegraphics[width=4.5in,height=4.5in]{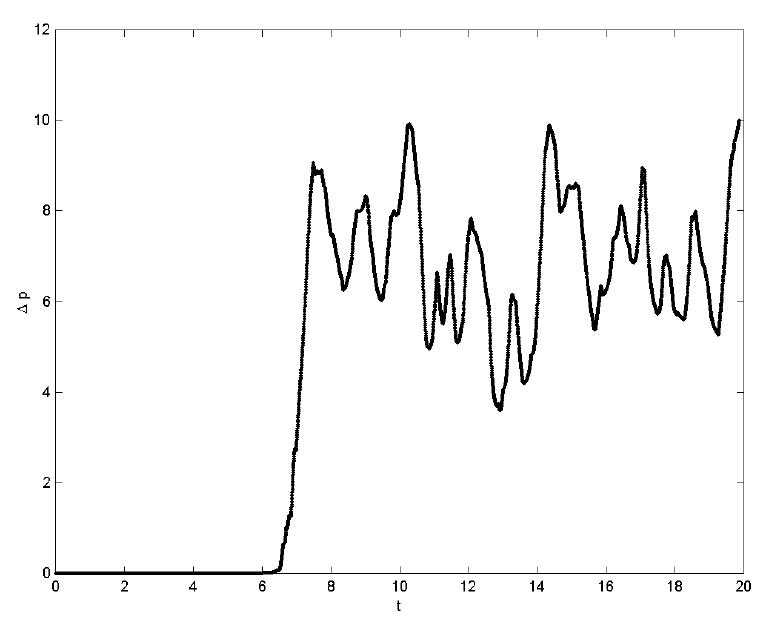}
\caption{Time series of the location deviations in Euclidean norm in the 2D case.}
\label{2DF1}
\end{figure}

\begin{figure}[ht] 
\centering
\includegraphics[width=4.5in,height=4.5in]{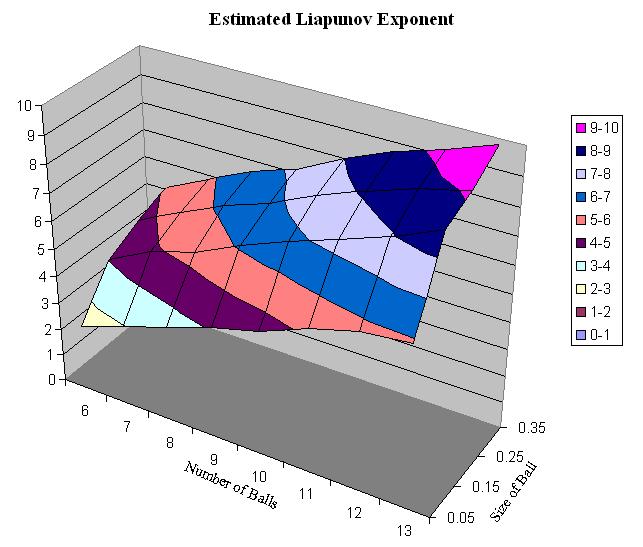}
\caption{The final estimated Liapunov exponent as a function of the size and the 
number of the disks.}
\label{2DF2}
\end{figure}

\begin{figure}[h] %[t]
\centering
\subfigure[$t=2.59$]{\includegraphics[width=2.3in,height=1.15in]{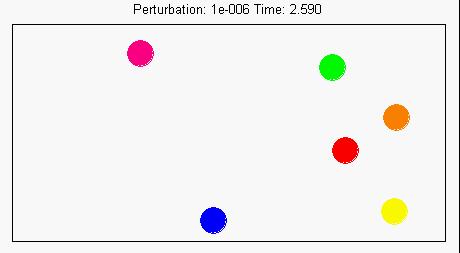}}
\subfigure[$t=3.286$]{\includegraphics[width=2.3in,height=1.15in]{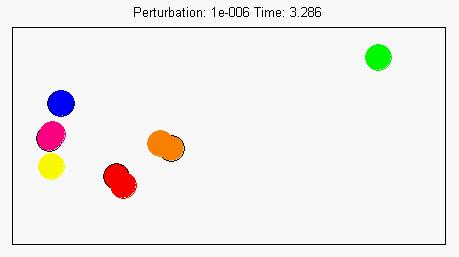}}
\subfigure[$t=3.42$]{\includegraphics[width=2.3in,height=1.15in]{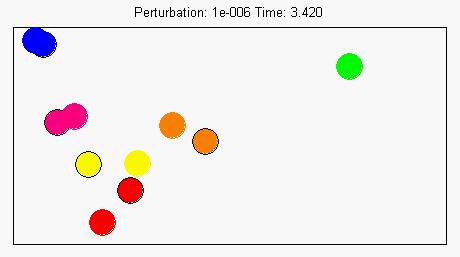}}
\subfigure[$t=3.954$]{\includegraphics[width=2.3in,height=1.15in]{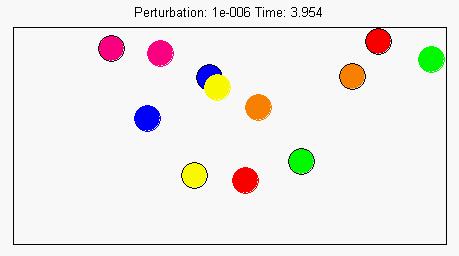}}
\caption{The evolution of the six disks and the evolution of the perturbed six disks. Since the perturbation 
size is $10^{-6}$, initially the unperturbed and the perturbed disks coincide almost completely. The radius of the disk 
is $0.25$, and the rectangle domain is $8\times 4$.}
\label{pan1}
\end{figure}

\begin{figure}[h] %[t]
\centering
\subfigure[$t=3.608$]{\includegraphics[width=2.3in,height=1.15in]{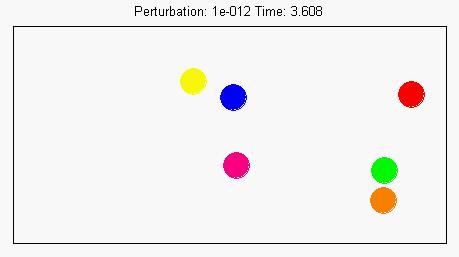}}
\subfigure[$t=4.046$]{\includegraphics[width=2.3in,height=1.15in]{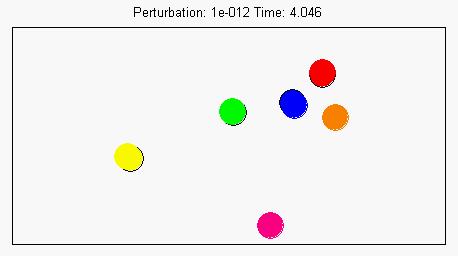}}
\subfigure[$t=4.184$]{\includegraphics[width=2.3in,height=1.15in]{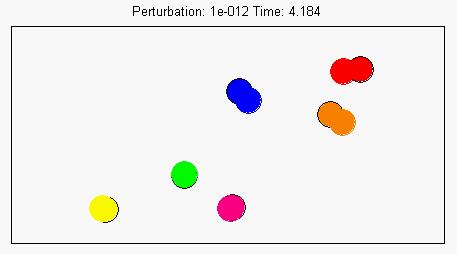}}
\subfigure[$t=5.248$]{\includegraphics[width=2.3in,height=1.15in]{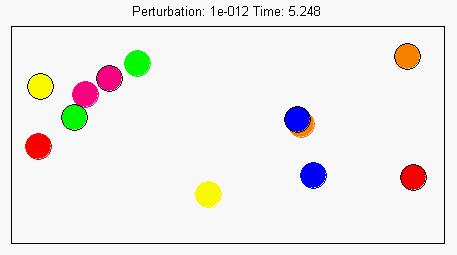}}
\caption{The same setup as in Figure \ref{pan1} except that the perturbation 
size is $10^{-12}$}
\label{pan2}
\end{figure}

The main setup is as follows: Particles move in the interval [$0,1$], particles collide with each other and 
the walls (the two boundaries $0$ and $1$ of the interval [$0,1$]) elastically. In particular, after they 
collide, two particles exchange their speeds to maintain the conservation of momentum  assuming all 
particles have the same mass; on the other hand, after it collides with a wall, a particle just reverses 
its velocity. 

\subsubsection{Reversibility}

In Figure \ref{1DF1}, initially all the particles are randomly positioned in the half interval [$0,0.5$],
and the initial velocities of the particles are randomly chosen from [$-0.5,0.5$]. After each collision, 
a random perturbation of magnitude less than $10^{-6}$ is introduced to the velocities, while no 
perturbation is introduced to the positions of the particles. Special care needs to be taken when a 
speed is small after a collision. As time increases, the particles will evolve into the full interval 
[$0,1$]. After $10,000$ collisions, the particles velocities are reversed and the time is reversed too. 
Then after each new collision, check if the particles are all inside the half interval [$0,0.5$]. 
The return probability $P_r$ is given by the quotient of the total number of collisions $N_r$ after 
which the particles are all inside the half interval [$0,0.5$] and $10,000$: 
\[
P_r = N_r / 10000 \ .
\]
For any fixed number of particles $N_p$, $10$ such numerical experiments are conducted, and the average 
return probability $\lag P_r \rag $ is ploted in Figure \ref{1DF1}. 

In Figure \ref{1DF2}, initially all the particles are randomly positioned in the full interval [$0,1$],
and the initial velocities of the particles are randomly chosen from [$-0.5,0.5$]. No random perturbation
is introduced. After each collision, check if the particles are all inside the half interval [$0,0.5$]. 
The numerical experiment ends after $10,000$ collisions. The return probability $P_r$ is defined the same as above. 
The average return probability $\lag P_r \rag $ of $10$ such numerical experiments  is ploted in Figure \ref{1DF2}. 

In Figure \ref{1DF3}, initially all the particles are evenly (with equal spacing) positioned in the full 
interval [$0,1$], and the initial velocities of the particles are randomly chosen from [$-0.5,0.5$]. 
After each collision, a random perturbation of magnitude less than $10^{-6}$ is introduced to the velocities, while no 
perturbation is introduced to the positions of the particles.  After each collision, check if the particles 
are all inside the half interval [$0,0.5$]. The numerical experiment ends after $50,000$ collisions. 
The return probability $P_r$ is defined the same as above (with $10,000$ replaced by $50,000$). 
The average return probability $\lag P_r \rag $ of $10$ such numerical experiments  is ploted in Figure \ref{1DF3}. 

From our preliminary numerical experiments, we conclude that
\begin{itemize}
\item Without perturbation, no matter how many particles are initially positioned in the half interval, and 
no matter how many total collisions, reversibility back to the half interval is always achieved as long as the 
computer accuracy is sufficient. This is due to the reversibility of the mathematical orbits. 
\item With perturbation, reversibility back to the half interval is possible for small number of particles 
($N_p < 16$). When the number of particle $N_p \geq 16$, extremely tiny perturbations (of magnitude less than $10^{-6}$) 
will lead to irreversibility. This is due to the sensitive dependence on initial data of the orbits in the phase space.
Take $N_p = 16$ as the example, in this case, the phase space has $32$ dimensions coordinated by the locations 
and velocities of the $16$ particles. When the $16$ particles are positioned in the half interval [$0,0.5$], and 
their initial velocities are chosen, a point in the $32$-dimensional phase space is chosen, from which a mathematical 
orbit exists. This rigorous mathematical orbit is reversible. But such a mathematical orbit has sensitive dependence 
on initial data. Even a tiny perturbation can lead to another diverging orbit. This causes the irreversibility 
of the numerical orbit when tiny random perturbations are introduced. We believe that the specifics of the 
perturbations and the specific ways in which they are introduced are important to irreversibility. What is crucial 
for irreversibility is the existence of perturbation. 
\end{itemize}

In Figure \ref{1DF4}, initially all the particles are randomly positioned in the half interval [$0,0.5$],
and the initial velocities of the particles are randomly chosen from [$-0.5,0.5$]. After each collision, 
a random perturbation of magnitude less than $10^{-6}$ is introduced to the velocities, while no 
perturbation is introduced to the positions of the particles. Special care needs to be taken when a 
speed is small after a collision. As time increases, the particles will evolve into the full interval 
[$0,1$]. After different relaxation time ($10000$, $20000$, $30000$), the particles velocities 
are reversed and the time is reversed too. Then the particles are tracked the same amount of time in the 
reverse time, at the end check if the particles are all inside the half interval [$0,0.5$]. The new return 
probability $P_{rv}$ (or more precisely reverse probability) is defined by the quotient of the number of experiments
with successful reverse $N_{rv}$ and the total number of experiments $N$:
\[
P_{rv} = N_{rv} / N \ .
\]
Due to the size of the computation, we usually run the number of experiments $N$ around $50$ to a few hundreds.

\subsubsection{Chaotic Dynamics}

First we look at the time series plots for two particles case ($N_p = 2$). In Figure \ref{fts}, for four different 
initial conditions, the location time series of the right particle are plotted. These time series look chaotic. 
But such chaotic dynamics may not be in the chaos in the classical sense. Bear in mind that at a collision, the 
velocity has a discontinuity. Next we explore further the chaotic dynamics along the line of Liapunov exponents. 

If we introduce small perturbations only on the initial locations of the particles, this will cause the perturbed 
collisions and the unperturbed collisions happen at a little different time, then the velocity deviation can be 
order one during a small time interval inbetween the perturbed and the unperturbed collisions. On the other hand, 
such a velocity deviation does not contribute substantially to the location deviation. So the location deviation 
maintains the same order as the initial location perturbation. Our numerical experiments verified this. 
Therefore, the main contribution for irreversibility comes from velocity perturbations.

If we introduce small perturbations only on the initial velocities (or both on the initial velocities and on the 
initial locations) of the particles, then the initial velocity
deviation will generates a location deviation which is linear in time, with speed given by the initial velocity
deviation. Near collisions, the location deviation will diverge from this main line, resulting in a rough 
oscillating continous non-differentiable location deviation function in time. The velocity deviation mainly 
follows the horizontal line in time given by the initial velocity deviation. Near collisions, it will just to 
other horizontal lines of order one away from the horizontal axis. Overall, the velocity deviation is a 
piecewise continuous function in time, with discontinuities at the the collisions, see Figures \ref{fvd1}-\ref{fvd2}.
For $N_p=2$, there are $5$ horizontal lines for velocity deviation (Figure \ref{fvd1}). The main line is the 
one very close to the horizontal line, given by the initial velocity deviation of order $10^{-6}$:
\[
\sqrt{(v'_1 - v_1)^2 + (v'_2 - v_2)^2},
\]
where ($v_1, v_2$) and ($v'_1, v'_2$) are the unperturbed and perturbed velocities of the two particles. 
The other four lines are near:
\[
2 v_1, \quad 2 v_2, \quad \sqrt{2} |v_1 + v_2|, \quad \sqrt{2} |v_1 - v_2| .
\]
For $N_p=3$, there are $10$ horizontal lines for velocity deviation. For $N_p=4$, there are $17$ horizontal 
lines for velocity deviation.  The location deviations are shown in Figures \ref{fld1}-\ref{fld2}.
The slope of the main line in the location deviation is given by the main line of the velocity deviation, which is 
very close to the horizontal line. The scattered points below the the main line in the location deviation 
come from the other lines of the velocity deviation, in lieu of collisions. Notice that, all the scattered points 
are below the main line in the location deviation. This can be explained from a closer look at the collision. 
Let $v_1$ and $v_2$ be the velocities of the unperturbed particles, $v_1+\hv_1$ and $v_2+\hv_2$ be the velocities 
of the perturbed particles. Without loss of generality, we assume that the unperturbed particles collide earlier.
Then there are essentially two possibilities as shown in Figure \ref{2CD}. In the case of Figure \ref{2CD}(a), 
right after the collision, both particle 1 and both particle 2's location deviations will decrease. In the case 
of Figure \ref{2CD}(b), right after the collision, particle 1 (the circle)'s location deviations will increase, 
while particle 2 (the square)'s location deviations will decrease. More specifically, right after the collision,
particle 1's velocity deviation is
\[
v_2 - (v_1 + \hv_1),
\]
while particle 2's velocity deviation is
\[
v_1 - (v_2 + \hv_2).
\]
Therefore, particle 1's velocity has an extra deviation (due to the collision) of 
\[
v_2 - v_1 ,
\]
while particle 2's velocity has an extra deviation of
\[
v_1 - v_2 .
\]
Thus the extra gain (due to the collision) of particle 1's location deviation is the same with the extra loss of 
particle 2's location deviation. In term of $\ell_1$-norm (sum of every particle's location deviation), in the case 
of Figure \ref{2CD}(b), the total location deviation of particles 1 and 2 is unchanged in lieu of the collision. 
On the other hand, in term of $\ell_2$-norm (Euclidean norm), the total location deviation may increase or decrease.
So far, we never observed the `increase' case. If a particle collides with a boundary, it 
is easy to see that the particle's location deviation will decrease right after the earlier collision. 

Scattered points below the main line in the location deviation always appear when the experiment time is long 
enough, see Figure \ref{fld3} for the two particles case. If we increase the size of the initial perturbation,
scattered points appear earlier. In Figures \ref{fld4} and \ref{fld5}, we use the initial perturbation size of 
$10^{-4}$ and the experiment time of $5000$ and $10000$, we can see a global picture in time of the location deviation
for the two particles case. Bear in mind that the blue region represents a continuous non-differentiable function 
in time. 

In conclusion, what we have observed is that the particle dynamics has sensitive dependence upon initial data,
but instead of ``exponential'' sensitive dependence like for traditional chaos, we observed here ``linear'' 
sensitive dependence; that is, the particle location deviation induced by an initial perturbation grows linearly 
in time,  instead of exponentially in time. It is such a linear growth that leads to irreversibility. 

\subsection{2-Dimensional Dynamics}

The main setup is now as follows: Circular disks of radius $r$ (instead of points in 1D case) move in the 
rectangle (of size $8 \times 4$), the elastic collision principle is applied to the disks colliding with each other 
and with the walls of the rectangle. The locations and velocities of the disks are identified by those of 
their centers. No rotational effect is considered. 

\subsubsection{Reversibility}

Limited numerical experiments indicates that the return probability $P_r$ (defined above) from the half to the whole 
rectangle follows the law of $P_r = 2^{-N}$ where $N$ is the number of disks; while in the 1D case, the 
return probability does not exactly follow this law due to particles blocking each other. 

\subsubsection{Chaotic Dynamics}

In the 2D case, the dynamics of the disks is completely chaotic. 

First (Figure  \ref{2DF1}) we choose the number of disks $N=6$, and the radius of each disk $r=0.05$. 
The initial velocities of 
the disks are randomly chosen with speed bounded by $10$. The initial locations of the disks are also 
randomly chosen within the rectangle. By adding a perturbation of size $10^{-15}$ to the initial locations, 
we observe the location deviations of the disks in time. Figure \ref{2DF1} shows the time series of the 
location deviations in Euclidean norm.

Next we compute the Liapunov exponent. For a perturbation of size $10^{-15}$ to the initial locations, we 
stop the computation when the Euclidean norm of the location deviations reaches $10^{-3}$. The time step 
of the computation is $0.002$. For each of the last $100$ time steps, we compute the quantity
\[
\frac{1}{\Dl t} \ln \frac{\Dl p (t +\Dl t)}{\Dl p (t)},
\]
where $\Dl p$ represents the Euclidean norm of the location deviations, and $\Dl t$ represents the time step.
Then we average these $100$ quantities to get an estimated Liapunov exponent for this numerical experiment. 
Finally, we average these estimated Liapunov exponents for $200$ such numerical experiments to obtain the final 
estimated Liapunov exponent. The dependence of the final estimated Liapunov exponent upon the size and the 
number of the disks is shown in Figure \ref{2DF2}.

Finally the slide shows on the dynamics of six disks are shown in Figures \ref{pan1} and \ref{pan2}.

\section{Epilogue}

We are convinced that the three ingredients mentioned in the 
section on Theory are enough to explain the mystery of the arrow of time. The level of physical rigor or 
mathematical rigor  that can be reached for this theory is unclear. The numerical simulations above 
(especially in the 2D case) offer positive support to our Theory. 

A huge amount of writings on the arrow of time exists in the literature, and we made no attempt to 
scan through them.

\end{document}